\documentclass[twocolumn,showpacs,preprintnumbers,amsmath,amssymb,nofootinbib]{revtex4}

\usepackage{natbib}
\usepackage{graphicx}
\begin{document}
\renewcommand{\v}[1]{\mathbf{#1}}
\def\x{{\mathbf{\hat{x}}}}
\def\y{{\mathbf{\hat{y}}}}
\def\z{{\mathbf{\hat{z}}}}

\def\sp{{\mathbf{\hat{s}}_p}}
\def\n0{{\mathbf{\hat{n}}_0}}
\def\Bs{{\mathbf{\beta}_s}}
\def\Bp{{\mathbf{\beta}_o}}

\def\doo{{\mathbf{\hat{d}}}}
\def\d0{{\mathbf{\hat{d}}_0}}
\def\dsd{{\Delta_s\mathbf{d}}}
\def\dod{{\Delta_o\mathbf{d}}}
\def\dsod{{\Delta_{so}\mathbf{d}}}
\def\Dd{{\Delta\mathbf{d}}}
\def\ro{{\mathbf{\hat{r}}}}
\def\r0{{\mathbf{\hat{r}}_0}}
\def\rc{{\mathbf{r_c}}}
\def\dsr{{\Delta_s\mathbf{ r}}}
\def\dor{{\Delta_o\mathbf{ r}}}
\def\dsor{{\Delta_{so}\mathbf{r}}}
\def\Dr{{\Delta\mathbf{r}}}
\def\mo{{\mathbf{\hat{m}}}}
\def\m0{{\mathbf{\hat{m}}_0}}
\def\dsm{{\Delta_s\mathbf{ m}}}
\def\dom{{\Delta_o\mathbf{ m}}}
\def\dsom{{\Delta_{so}\mathbf{ m}}}
\def\Dm{{\Delta\mathbf{ m}}}
\def\ho{{\mathbf{\hat{h}}}}
\def\h0{h_0}

\def\Ai{[3\cos^2\theta+1]^\frac{1}{2}}
\def\epsilon{\varepsilon}
\def\eO{\epsilon_s}

\def\ba{\begin{eqnarray}}
\def\ea{\end{eqnarray}}

\title{The Effects of Finite Emission Height in Precision Pulsar Timing}

\author{Caroline D'Angelo}
\altaffiliation{Now at the Max-Planck-Institut f\"ur Astrophysik, Karl-Schwarzschild-Str. 1
 Garching, Germany, 85741}
\affiliation{Department of Astronomy and Astrophysics, University of Toronto,
    50 St. George Street, Toronto, ON M5S 3H4, Canada}
\email{dangelo@astro.utoronto.ca}
\author{Roman R. Rafikov}
\email{rrr@cita.utoronto.ca}
\affiliation{CITA, McLennan Physics Labs, 60 St. George St., 
University of Toronto, Toronto, ON M5S 3H8, Canada}

\date{\today}

\begin{abstract}
Precision timing is the key ingredient of ongoing
pulsar-based gravitational wave searches and 
tests of general relativity using binary pulsars.
The conventional approach to timing explicitly 
assumes that the radio emitting region 
is located at the center of the pulsar, while polarimetric 
observations suggest that  radio emission is 
in fact produced at altitudes ranging 
from tens to thousands of kilometers above the neutron star 
surface. Here we present a calculation of the effects of 
finite emission height on the timing of binary pulsars 
using a simple model for the emitting region geometry. 
Finite height of emission $h$ changes the propagation 
path of radio photons through the binary and gives rise 
to a large spin velocity of the emission region 
co-rotating with the neutron star. Under favorable 
conditions these two effects 
introduce corrections to the conventional time delays 
at the microsecond level (for a millisecond pulsar 
in a double neutron star 
binary with a period of several hours and assuming 
$h=100$ km). Exploiting the dependence of the 
emission height on frequency (radius-to-frequency mapping) 
and using multi-frequency observations one should be 
able to detect these timing corrections even though they 
are formally degenerate with conventional time delays. 
Although even in the most accurately timed systems 
the magnitude of the finite emission height effects 
is currently somewhat below timing precision, 
longer-term observations and future facilities 
like SKA will make measurement of these effects possible,
providing an independent check 
of existing emission height estimates.
\end{abstract}


\maketitle
\section{Introduction}
\label{sect:intro}

Since their discovery radio pulsars have been recognized as 
precise time-keepers. This unique property has been extensively 
exploited to test  general relativity 
since 1975, when Hulse and Taylor discovered the first double 
neutron star (DNS) system PSR 1913+16 \cite{1975ApJ...195L..51H}. 
Thirty years later, pulsar surveys have found seven other DNS systems 
(among them a double pulsar system J0737-3039, in which both 
neutron stars are seen as radio pulsars, see 
\cite{2004Sci...303.1153L}) and many white dwarf-neutron 
star (WD-NS) binaries. These systems have been used not 
only to test a number of general relativistic effects but 
also to set limits on the stochastic gravitational wave 
background \cite{2006astro.ph..9013J}.

Precise monitoring of the times of arrival (TOAs) of pulses 
on Earth is the key to success in this game. Although 
timing accuracy at the several $\mu$s level has been 
routinely achieved in a number of systems, only three WD-NS
binary pulsars -- J0737-4715, J1713+0747, J1909-3744 
\cite{2006MNRAS.369.1502H,2006astro.ph..9013J} -- 
have been timed to $0.1-0.3~\mu$s precision. 
At the same time, current instruments have not yet reached the 
level of intrinsic uncertainty present in pulsars 
(which will ultimately set a limit on their precision, 
see \cite{2004NewAR..48.1413C}). 
Ambitious new projects like the Parkes 
Pulsar Timing Array (PPTA), which aims to achieve daily 
timing residuals $\lesssim 0.1~\mu$s for more than ten 
millisecond pulsars \cite{2006IAUJD..16E..66M}, 
and proposed facilities like the Square Kilometer Array 
\cite[SKA;][]{2004NewAR..48.1413C}, demand an understanding  
of the systematic timing uncertainties at the level of 
$1$ ns \cite{2006MNRAS.tmp.1152E}.
With these new instruments, effects previously hidden in 
error will become observable and some of them may have not 
been accounted for by the current timing models. 

The conventional physical model for interpreting timing data in 
terms of the binary system's parameters, which was advanced by 
\citet[hereafter DD]{1985JAF....25...21D} and extended by 
\citet[hereafter DT]{1992PhRvD..45.1840D}, 
treats the binary as two point masses in orbit around each other
and assumes the radio emission region to be located at the very 
center of the radio pulsar. However, theoretical models of 
pulsar radio emission, detailed analysis of pulse structure 
and polarimetry of isolated pulsars suggest that their 
radio emission is in fact produced high 
in the magnetosphere. In slow pulsars it occurs at 
typical distances of order several hundred (or thousand in some 
cases) kilometers from the neutron star \cite{1991ApJ...370..643B,
2001ApJ...555...31G,2004ApJ...607..939D,2004A&A...421..215M}. 
For millisecond pulsars, which are more 
appropriate for precision timing, the typical height of the emission 
region $h$ is estimated to be several tens of kilometers 
\cite{1993A&A...273..563G,2003A&A...397..969K}, 
although at low frequencies $h$ may 
be as high as $\approx 100$ km.

A finite emission height has  twofold effect on pulsar timing.
First, the displacement of the emission region from the 
neutron star center affects photon propagation through
the curved space-time within the binary. Second, 
a high-altitude emission region of a rapidly spinning 
millisecond pulsar must be moving at a rather 
large speed (in some cases reaching $0.2c$ 
\cite{2005ApJ...628..923G}, where $c$ is the speed of light),
leading to severe aberration of the pulsar radio beam. 
For a given spin period of a pulsar $P_p=2\pi/\Omega_p$ 
the upper limit on the emission height $h$ is set by 
the size of the light cylinder $r_{LC}=c/\Omega_p\approx 
500(P_p/10~\mbox{ms})$ km, while the lower limit on 
$h$ is set by the neutron star radius $R_{NS}\approx 10$ km. 
The latter 
also sets a lower limit of $\Omega_p R_{NS}\sin\alpha\approx
0.02c(P_p/10~\mbox{ms})^{-1}\sin\alpha$ on the 
rotational velocity of the
emitting region, where $\alpha$ is the angle between the
magnetic and spin axes of the pulsar. 
 
Given the rapidly improving accuracy of pulsar timing and the 
increasing demand to understand systematic timing effects 
at the nanosecond level, which are driven by ongoing and 
future projects, the aforementioned effects of a non-zero emission 
height merit closer inspection. The aim of this paper is to 
extend the DT model of pulsar timing to the case where the 
emission comes from a region high in the 
magnetosphere, and determine whether this physical change 
produces any measurable results. 

An outline of the paper runs as follows. In sec. \ref{sec:dipole} 
we describe the geometry of the emission in a dipolar 
magnetic field. In sec. \ref{sec:timing}, we consider different 
timing effects and compute corrections that result from considering a finite emission height. We 
discuss the measurability of these corrections in 
sec. \ref{sec:measurability} and limitations and possible 
refinements of the model in sec. \ref{sect:discussion}.

\section{Geometrical Model of Radio Emission}
\label{sec:dipole}

\begin{figure}
\includegraphics[width=\hsize]{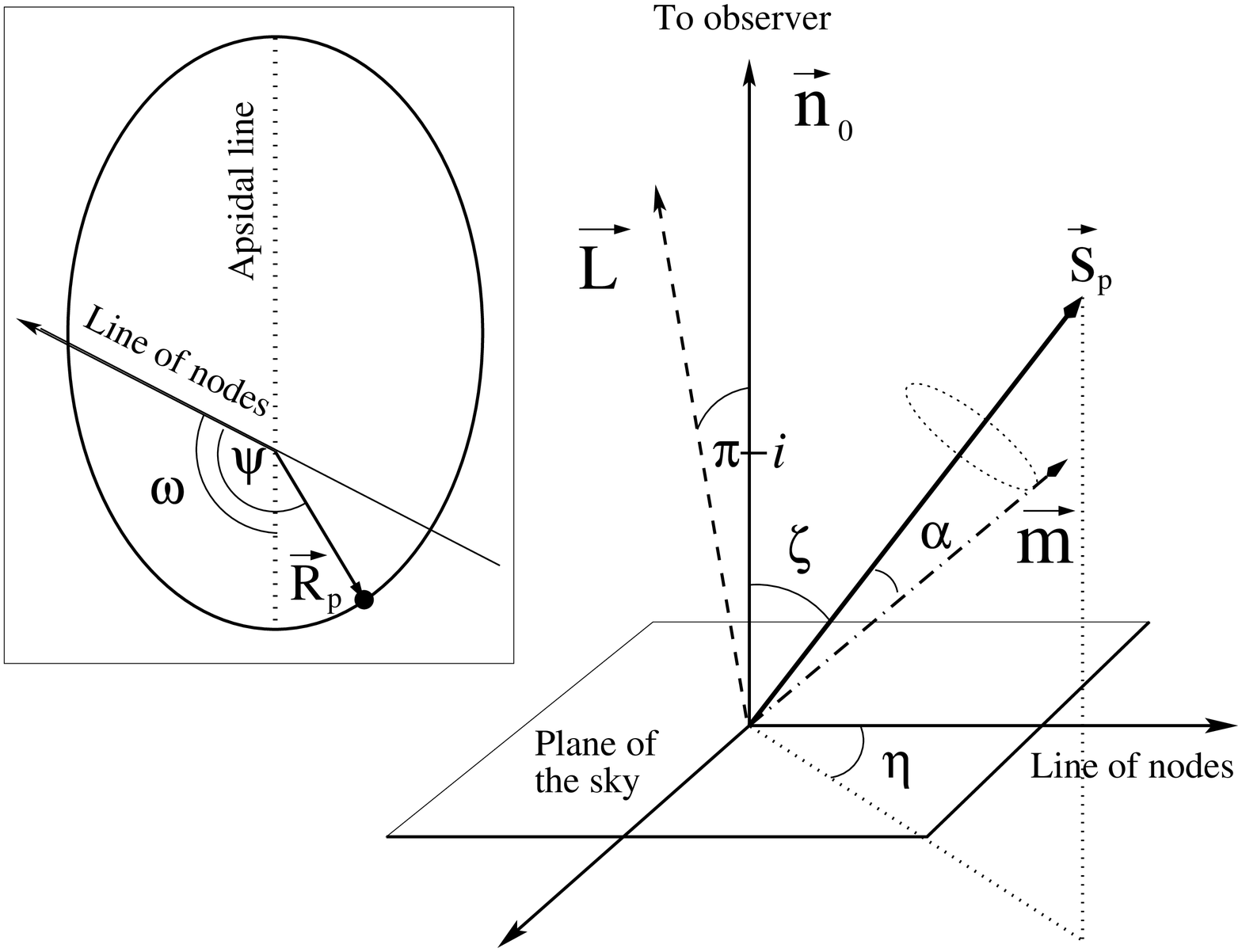}
\caption{Orbital configuration and spin orientation of the pulsar.
Here $\sp$ is the spin axis for the neutron star, $\n0$ is the 
direction to the observer, $\mo$ is the magnetic field axis,
${\bf R}_p$ is the position vector of the pulsar. We use the orientation
notation of DT so that the orbital angular momentum vector ${\bf L}$ 
constitutes angle $i$ with $-\n0$. See text for 
details.}\label{fig:orientation}
\end{figure}

We consider a binary system with a semimajor axis $a$, 
eccentricity $e$, and inclination $i$ with respect to the 
observer's line of sight. The instantaneous position 
of the pulsar is characterized by the true anomaly 
$\psi$ measured from the ascending node of the pulsar 
orbit, and $\omega$ is the longitude of periastron, see
Figure \ref{fig:orientation}. We use ${\bf R}_p$ and 
${\bf R}_c$ for the position vectors of the pulsar and its companion
with respect to the barycenter of the binary. In making numerical 
estimates we will adopt a fiducial model 
of the pulsar binary (motivated by the properties of the 
double pulsar J0737-3039, see \cite{2004Sci...303.1153L}) 
which consists of two neutron stars
($M_p=M_c=1.4~M_\odot$ are the assumed pulsar and companion 
masses) in a compact orbit with a semimajor axis $a=10^{11}$
cm (orbital period $P_b=2.9$ hrs). We take pulsar spin 
period to be $P_p=10$ ms and assume the height 
of the emission region to be $h_{7}=10^7$ cm. 

\begin{figure*}
\includegraphics[width=\hsize]{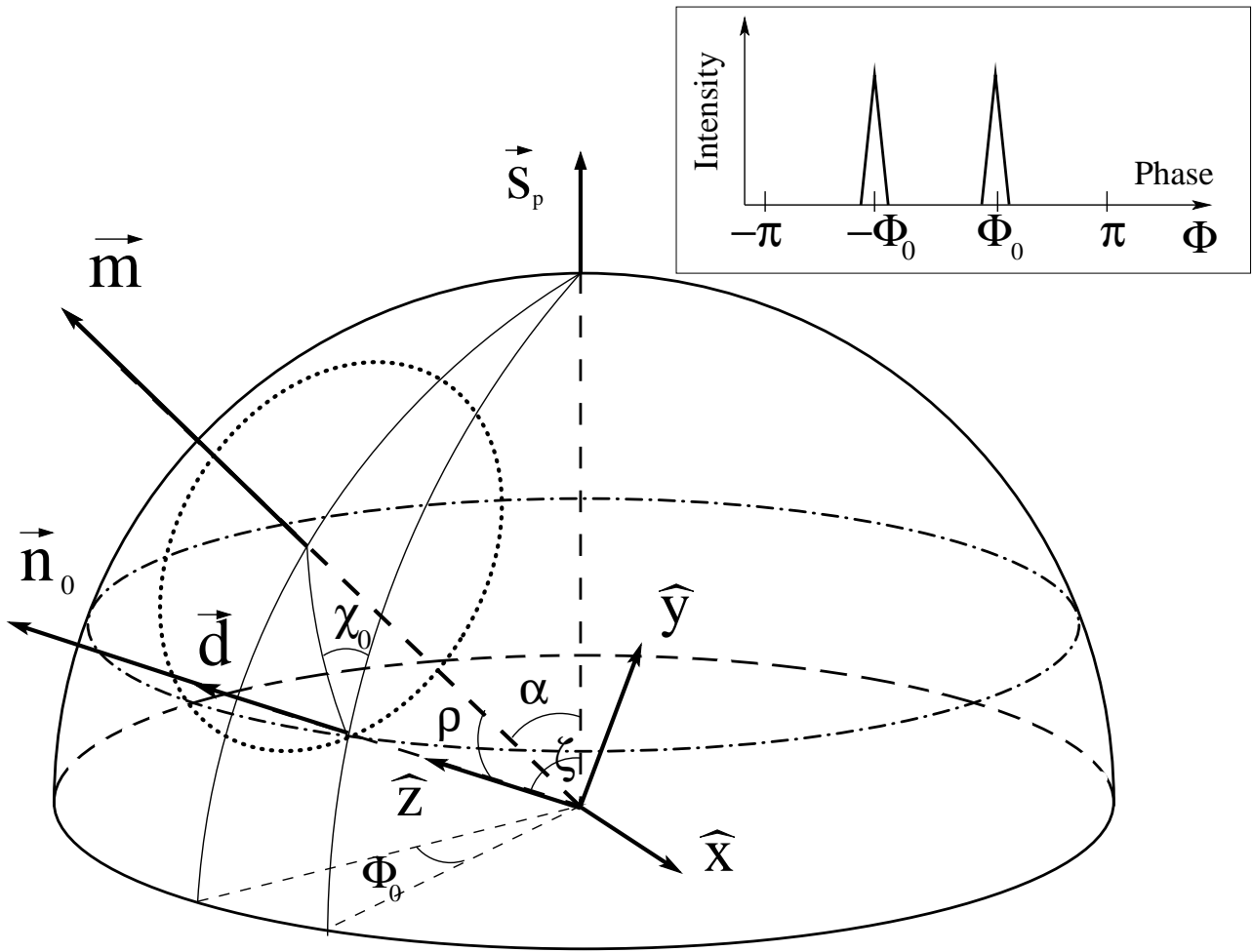}
\caption{Model of the emission cone geometry of the pulsar. 
Here we show the direction of emission $\doo$ 
when $\doo=\n0$ and the pulse is observed. Note that 
at the moment of emission 
$\mo$ has two values, which correspond to the 
two crossings of the emission cone by the observer's 
line of sight. The coordinate system $(\x,\y,\z)$ is also
displayed.
}\label{fig:setup}
\end{figure*}

\begin{figure*}
\begin{center}
\includegraphics[width=.5\hsize]{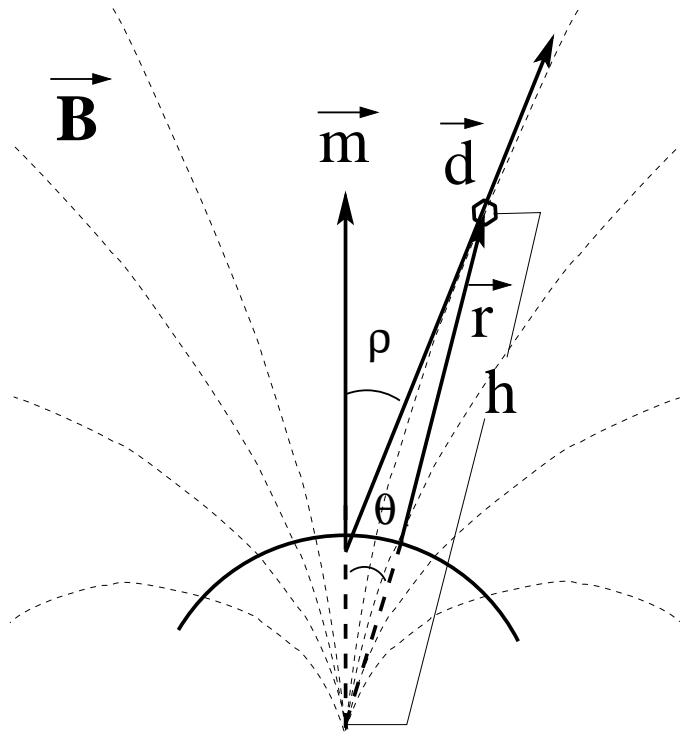}
\caption{An illustration of the dipole model used in our 
analysis. The field lines emanate from the surface of the 
star about the magnetic field axis, $\mo$, and the emission 
is produced at the point h$\ro$, in the direction $\doo$, 
which is tangent to the field line.}\label{fig:dipole}
\end{center}
\end{figure*}

To illustrate our calculations 
we adopt a specific model for the geometry of pulsar 
radio emission closely related to the conventional rotating 
vector model \cite[RVM;][]{1969ApL.....3..225R}. 
In the comoving coordinate system of the pulsar we assume  
neutron star magnetic field to have a dipolar geometry
(which should be a good approximation at the distances where
radio emission is produced), with the magnetic axis $\mo$ ($|\mo|=1$) 
making an angle $\alpha$ with the pulsar spin axis 
$\sp$ ($|\sp|=1$), see Figure \ref{fig:setup}. 
Vector $\sp$ makes an angle $\zeta$ with the unit 
vector $\n0$ from the binary system to observer at 
Earth ($\sp\cdot\n0=\cos\zeta$), while the projection
of $\sp$ on the plane of the sky makes an angle $\eta$
with the ascending node of the binary orbit, see 
Figure \ref{fig:orientation}.
While the vector $\sp$ is fixed, the orientation of  
$\mo$ changes in time because the pulsar rotation 
always maintains the relation
\ba
\mo\cdot\sp=\cos\alpha.
\label{eq:mag_spin}
\ea  
The position of the emitting region with respect 
to the pulsar center is $h\ro$, where $|\ro|=1$ and 
$h$ is the emission height, see Figure \ref{fig:dipole}. 
Emission is assumed to occur strictly
along the local ${\bf B}$ field since it is likely produced
by the highly relativistic plasma streaming along the 
open field lines, so that the 
direction of emission $\doo$ ($|\doo|=1$) is related to $\ro$ and 
$\mo$ as
\ba
\doo = \frac{3\cos\theta\ro - \mo}{(3\cos^2\theta+1)^{1/2}},
\label{eq:d_through_r_m}
\ea
where $\theta$ is the angle between $\ro$ and $\mo$, i.e.
\ba
\mo\cdot\ro=\cos\theta.
\label{eq:m_r}
\ea 
Vector $\doo$ makes an angle $\rho$ with the magnetic axis $\mo$
(see Figure \ref{fig:dipole}) such that 
[see eq. (\ref{eq:d_through_r_m})]
\ba
\cos\rho=\frac{3\cos^2\theta - 1}{(3\cos^2\theta+1)^{1/2}}.
\label{eq:rho}
\ea
 
We will be considering emission by all sources at the same 
height $h$ belonging to a single flux surface  
for which the condition (\ref{eq:m_r})
with a given value of $\theta$ is fulfilled. 
Thus, $h$ and $\theta$ are fixed for this calculation
(but see sec. \ref{sect:var_height} for the 
consequences of relaxing this assumption). All such sources 
form a circular emission region around vector $\mo$, 
which gives rise to an emission cone with an opening angle 
$2\rho$ rigidly connected to (and co-axial with) the 
magnetic axis. When the emission cone crosses our line of sight
as a result of pulsar rotation (that is, for some point $h\ro$ 
in the emission region, $\n0$ becomes parallel 
to $\doo$) a pulse of emission
is sent to the observer on Earth. There are two 
emission episodes per spin period of pulsar corresponding to 
the two emission cone crossings by $\n0$, separated by 
$2\Phi_0$ in pulse phase (see Figure \ref{fig:setup}), where
\ba
\cos\Phi_0=\frac{\cos\rho-\cos\zeta\cos\alpha}
{\sin\zeta\sin\alpha}.
\label{eq:Phi_0}
\ea
By measuring the TOAs of these pulses
one can obtain information about the motion of the 
pulsar and the propagation effects inside the binary. 

The emission cone crosses our line of sight and makes the pulsar 
observable only if $|\zeta-\alpha|<\rho$. 
In the absence of aberration caused by the motion of the
emission region (i.e. when pulsar is not spinning and is at rest
with respect to observer on Earth), favorable conditions
for the detection of this conal emission,
\ba
\doo=\n0,
\label{eq:master_eq}
\ea
can be achieved only for some specific orientations of 
vectors $\ro$ and $\mo$, which we call $\r0$ and $\m0$.
To find these zeroth order (or static) values of 
$\ro$ and $\mo$ we form a stationary orthogonal 
basis using $\sp$ and $\n0$:
\ba
(\x, \y, \z)  = \left(\frac{\sp\times\n0}{|\sp\times\n0|}, 
\frac{\sp - \cos\zeta\n0}{|\sp\times\n0|}, \n0\right).
\label{eq:basis}
\ea
Then, using equations (\ref{eq:mag_spin}), 
(\ref{eq:d_through_r_m}), (\ref{eq:m_r}),
and (\ref{eq:master_eq}) one can easily find that
\begin{eqnarray}
\m0 &=& \pm\sin\Phi_0\sin\alpha~\x+
\frac{\cos\alpha-\cos\zeta\cos\rho}{\sin{\zeta}}\y\\
\nonumber&+&\cos\rho~\z,
\label{eq:m0}\\
\r0 &=& \pm
\frac{\sin\Phi_0\sin\alpha}{3\cos\theta}\x
+\frac{\cos\alpha-\cos\zeta\cos\rho}
{3\cos\theta\sin\zeta}\y\\
\nonumber &+&\frac{\cos\rho+\Ai}{3\cos\theta}\z,
\label{eq:r0}
\end{eqnarray}
where $\sin\Phi_0$ can be found from equation 
(\ref{eq:Phi_0}).
The $\pm$ ambiguity in the $x$-components of $\m0$ 
and $\r0$ corresponds to the two crossings of the 
emission cone by $\n0$ which results 
in two possible orientations of $\m0$ and $\r0$, 
see Figure \ref{fig:setup}. The upper and lower signs are for 
the leading and trailing pulses, respectively (cone crossings).

Slow pulsars typically have narrow beams meaning that
$\rho\ll 1$, $\Phi_0\ll1$, and 
$\theta\approx (2/3)^{1/2}\rho\ll 1$. In this case it 
follows from equations (\ref{eq:m0}) and (\ref{eq:r0})
that ${\bf \hat m}_{0,x}, {\bf \hat m}_{0,y}, 
{\bf \hat r}_{0,x}, {\bf \hat r}_{0,y}\sim \theta\ll 1$ 
(remember that $|\zeta-\alpha|<\rho$) so that both 
$\m0$ and $\r0$ are almost aligned with 
$\n0$. On the other hand, millisecond pulsars 
(the objects most suitable for precise timing), have wide beams 
with $\rho, \Phi_0, \theta\sim 1$, so that their $\m0$ 
and $\r0$ are generally quite mis-aligned with $\n0$.

When the emission region is moving, the concomitant 
aberration changes the emission direction $\doo$ 
as seen by an observer at rest, so that the radio pulses 
are detected only at $\ro$ and $\mo$, which are somewhat different 
from $\r0$ and $\m0$. The conventional timing model (DD)
assumes that the motion of the emission region 
comes only from the orbital motion of pulsar
with speed ${\bf v}_o=\Bp c$ (the proper motion of 
the binary as a whole is undetectable through timing). 
A finite height of emission introduces the spin velocity 
${\bf v}_s=\Bs c$, since the emission  
region co-rotates with the pulsar. In the case of 
millisecond pulsars this spin velocity is a 
considerable fraction of the speed of light 
\cite{2005ApJ...628..923G} even if emission takes
place at the surface of the neutron star 
($v_s\approx 0.2c$ for an orthogonal rotator with 
spin period $1$ ms and $h\sim 10$ km, which is 
much larger than $v_o\sim 10^{-3}c$).

In the general case of nonzero orbital and spin
velocities of the emitting region unit vectors
$\ro$ and $\mo$ can be written as
\begin{eqnarray}
\ro &=& \r0 + \dor+ \dsr  + \dsor,
\label{eq:ro}\\
\mo &=& \m0  + \dom + \dsm + \dsom,
\label{eq:mo}
\end{eqnarray}
where $\Delta_o$ denotes corrections due to 
${\bf v}_o$ (which have been previously calculated in the framework of 
DD timing model), $\Delta_s$ corresponds to the corrections 
due to ${\bf v}_s$ only, while $\Delta_{so}$ corresponds to 
the coupling between ${\bf v}_o$ and ${\bf v}_s$. The 
derivation of all these corrections is presented in Appendix
and the expressions for $\dor, \dsr, \dsor, \dom, \dsm, \dsom$
are given by equations (\ref{eq:dsm})-(\ref{eq:dsr}) and 
(\ref{eq:dom})-(\ref{eq:dsor}).
For the parameters of our fiducial pulsar binary model 
the typical relative magnitudes of the 
different correction terms listed in equations
(\ref{eq:ro}) and (\ref{eq:mo}) are 
\begin{eqnarray}
\epsilon_o &=& \frac{v_{o}}{c}\approx 10^{-3}a_{11}^{-1/2},
\label{eq:eps_orb} \\
\epsilon_s &=&  \frac{\Omega_p h}{c} = 
0.21~h_7\left(\frac{P_p}{10~\mbox{ms}}\right)^{-1},
\label{eq:eps_spin} \\
\epsilon_{so}& =& \epsilon_o\epsilon_s \approx 2\times 10^{-4}
h_7a_{11}^{-1/2}\left(\frac{P_p}{10~\mbox{ms}}\right)^{-1},
\label{eq:eps_so}
\end{eqnarray} 
correspondingly, where $h_{7}\equiv h/(10^{7}\mbox{cm})$ and 
$a_{11}\equiv a/(10^{11}\mbox{cm})$.

\section{Effect on timing}
\label{sec:timing}

We are now in a position to calculate 
the corrections to the time delays that occur in  
binary pulsar systems. These delays appear when one
relates the observer's proper TOA of a given radio 
pulse $\tau_a$ to the proper time of its emission in the 
comoving frame of the pulsar $T_e$. As demonstrated by DD 
this transformation can be done in several steps, 
going first from $\tau_a$ 
to the coordinate time of arrival $t_a$ (measured in 
a frame comoving with the binary barycenter), which 
is linked to the coordinate time of emission $t_e$  through
the two propagation delays -- the R\"omer ($\Delta_R$) and 
Shapiro ($\Delta_S$) delays (see eq. [4] of DD). 
In its turn, the coordinate time $t_e$ is directly related 
to the proper emission time in the pulsar frame $T_e$ 
via the Einstein delay $\Delta_E$, so that 
$t_e=T_e+\Delta_E$. Finally, the deviation 
of the proper time of emission $T_e$ from the purely 
periodic pattern in the pulsar frame is determined 
by the so-called aberration delay $\Delta_A$. 

The introduction of the nonzero emission height 
modifies some of these delays. The R\"omer and Shapiro
delays get affected because of the change in the 
path of the radio beam. The aberration delay gets modified
because finite height introduces spin motion of the
emission region (in addition to the orbital motion)
which affects aberration of the radio signal. At the same 
time, the Einstein delay is not affected by the finite 
emission height as $\Delta_E$ is simply the 
difference between the proper time and the coordinate 
time of pulsar, which is independent of the geometry 
and physics of emission.

When calculating the timing contributions we will neglect 
the effect of gravitational lensing by the companion which 
mainly affects the Shapiro delay and may be relevant only for
highly inclined binary systems \cite{1990A&A...232...62S,
1995MNRAS.274.1029D,2005ApJ...621L..41L,2006ApJ...641..438R}.
We will also neglect the retardation effect 
\cite{1999PhRvD..60l4002K,2006PhRvD..73f3003R} which can 
be easily accounted for afterwards.

\subsection{R\"omer Delay}
\label{sec:roemer}

We begin with the R\"omer delay $\Delta_R$, the variation of 
the photon travel time as the pulsar moves through its orbit. 
This delay can be expressed as:
\begin{equation}
\label{eq:roemer}
\Delta_R = -\frac{1}{c}\v{\hat{n}_0}\cdot{\bf R}_e,
\end{equation}
where ${\bf R}_e$ is the position of the emission region 
relative to the binary barycenter. The conventional DD model 
assumes ${\bf R}_e = {\bf R}_p$, while in our case 
${\bf R}_e={\bf R}_p+ h\ro$, which gives 
rise to a correction to the R\"omer delay $\Delta_{R,h}$.
We use equation (\ref{eq:ro}) to calculate $\Delta_{R,h}$, in which 
$\r0$ and $\dsr$ are constant and would not contribute 
to the timing signal, while $\dsor$ is small compared to
$\dor$ by $\sim \epsilon_s$ (see below). Thus, we only 
retain $\dor$ and find 
\begin{eqnarray}
\Delta_{R,h} &=& - 
\frac{(\Bp_\perp\cdot\m0)}{3\cos\theta}
\frac{h}{c}\nonumber\\
& =& -\frac{\h0}{3c\cos\theta}
\left[\pm(\Bp\cdot\x)\sin\Phi_0\sin\alpha  \right.\nonumber\\
&&\left.+(\Bp\cdot\y)\frac{\cos\alpha-\cos\zeta\cos\rho}{\sin\zeta}\right]
\label{eq:modroemer}
\end{eqnarray}
where for any vector $\v{A}$, we denote its component perpendicular to
$\n0$ as ${\bf A}_\perp \equiv {\bf A}- ({\bf A}\cdot\n0)\n0$. 
Note that the first term in the right-hand side of this 
expression has a sign ambiguity, which reflects the term's opposite
timing contribution to the leading and trailing components of the
pulse profile 
(corresponding to the two crossings of the same emission cone 
by vector $\doo$). This ambiguity  has the
effect of periodically changing the separation between the 
two components of the pulse profile on an orbital timescale, 
which makes this timing contribution similar to 
the so-called latitudinal time delay caused by orbital aberration 
\cite{1992PhRvD..45.1840D,2006ApJ...641..438R}. 
The second term in (\ref{eq:modroemer}) exhibits no such
sign ambiguity, meaning that it is the same 
for both pulse components, and manifests itself 
as an overall homogeneous shift of the pulse profile, 
similar to other conventional time delays. 

Assuming that all the angles specifying pulsar spin-magnetic 
orientation in (\ref{eq:modroemer}) are of order unity 
(typical for rapidly spinning millisecond 
pulsars) one finds that the typical amplitude of 
$\Delta_{R,h}$ is 
\ba
t_h=\frac{h}{c}\Bp\approx 0.3~\mu\mbox{s} ~h_7
\left(\Bp/10^{-3}\right),
\label{eq:t_h}
\ea
which is within the timing accuracy of the most stable pulsars.
For slower pulsars which typically have small 
$\theta\sim \rho$ and $\Phi_0$, the expression 
in brackets in (\ref{eq:modroemer}) would be 
of order $\theta\ll 1$, so that 
$\Delta_{R,h}\sim \theta t_h$, making this 
contribution difficult to measure.

One may ask whether it 
is legitimate to discard $\dsor$ contribution in 
(\ref{eq:modroemer}) when $\theta\ll 1$. We believe it is,
since the inclusion of $|\dsor|$ would add terms 
of relative significance $\sim\epsilon_s$. At the same time, 
angles $\rho$
and $\theta$ are ultimately related to the position
of the last open flux surface which touches the light 
cylinder at $r_{LC}=c/\Omega_p$. At the distance 
$h$ this flux surface is characterized by angle $\theta\sim 
(h/r_{LC})^{1/2}=\epsilon_s^{1/2}$. Thus, $|\dsor|$ 
contributions
are smaller than $\Delta_{R,h}\sim \theta t_h$ by 
$\sim \epsilon_s^{1/2}\ll 1$ 
and can be neglected.

\subsection{Shapiro Delay}
\label{ssect:Shapiro}

The Shapiro delay is caused by the curvature of space-time near 
the companion, which is only substantial when the pulse passes 
near the companion. It is  accurately measurable only in 
nearly edge-on systems. 
The conventional formula for the Shapiro delay is
\begin{equation}
\Delta_{S,0} = -\frac{R_g}{c}\log[{\v{\hat{n}_0}}
\cdot({\v{R}_p}-{\v{R_c}}) - |{\v{R}_p}-{\v{R}_c}|] + const,
\label{eq:Shap_conv}
\end{equation}
where $R_g\equiv 2GM_c/c^2$ is the Schwarzschild radius
of the companion.

A displacement of the emission region from 
the center of pulsar changes the propagation path of the
radio beam and affects the Shapiro delay. The modified
formula for the delay can be obtained by changing
$\v{R}_p$ in equation (\ref{eq:Shap_conv}) to 
${\bf R}_e={\bf R}_p+ h\ro$ and reads
\ba
\Delta_S &=& -\frac{R_g}{c}\log
\left[\n0\cdot(\v{R}_p+h\ro-\v{R}_c) \right.\nonumber\\
&-&\left.|\v{R}_p+h\ro-\v{R}_c|\right]
\nonumber\\
&\approx & -\frac{R_g}{c}\log
\left[(\v{R}_{p}-\v{R}_c)\cdot (\n0-{\bf \hat n}_R) \right.\nonumber\\
&+&\left. h\r0\cdot\left(\n0-{\bf \hat n}_R \right)\right],
\label{eq:shapiro}
\ea
where ${\bf \hat n}_R=(\v{R}_{b}-\v{R}_c)/|\v{R}_{b}-\v{R}_c|$,
and in the second line we have expanded the expression 
inside the logarithm in terms of $h/|\v{R}_{b}-\v{R}_c|$,
while also setting $\ro=\r0$. This can be rewritten as
\ba
&& \Delta_S =\Delta_{S,0}+\Delta_{S,h},
\label{eq:sgap_gen}\\ 
&& \Delta_{S,h}=-\frac{R_g}{c}
\frac{h\r0\cdot\left(\n0-{\bf \hat n}_R \right)}
{(\v{R}_{p}-\v{R}_c)\cdot (\n0-{\bf \hat n}_R)}.
\label{eq:shapiro3}
\ea

Far from conjunction the height related correction is 
very small: $\Delta_{S,h}\sim (h/a)(R_g/c)\lesssim 0.1$ ns
since $h/a\sim 10^{-4}$ and $R_g/c\sim 10~\mu$s even in DNS 
binaries. However, around conjunction
$(\v{R}_{b}-\v{R}_c)\cdot (\n0-{\bf \hat n}_R)
\sim b^2/|\v{R}_{b}-\v{R}_c|\ll a$, where 
$b\approx a\cos i$ is the minimum projected separation 
between the pulsar and its companion in the plane of 
the sky. At the same time, 
\ba
\r0\cdot\left(\n0- {\bf \hat n}_R\right)&=&\pm
\frac{\sin\Phi_0\sin\alpha}{3\cos\theta}
(\n0- {\bf \hat n}_R)_x
\label{eq:height_corr}\\
\nonumber&+&\frac{\cos\alpha-\cos\zeta\cos\rho}
{3\cos\theta\sin\zeta}(\n0- {\bf \hat n}_R)_y
\nonumber\\
&+&\frac{\cos\rho+\Ai}{3\cos\theta}
(\n0-{\bf \hat n}_R)_z,
\nonumber
\ea
and at conjunction $(\n0-{\bf \hat n}_R)_z
\sim b^2/|\v{R}_{b}-\v{R}_c|$,
while the first two terms in the right-hand side are of
order $\theta b/|\v{R}_{b}-\v{R}_c|$. Thus, for 
millisecond pulsars with $\rho, \theta\sim 1$ 
$\Delta_{S,h}$ goes up to $\sim (h/b)(R_g/c)$ 
at conjunction. This can be substantial
in a DNS binary: for $a=10^{11}$ cm, $h\sim 10^2$ km, 
and $i=89^\circ$ one has
$b\approx 17,000$ km so that $h/b\sim 10^{-2}$ and at 
conjunction $\Delta_{S,h}$ can reach $\sim 0.05~\mu$s.

We also remark that the sign ambiguity of the first 
term in equation (\ref{eq:height_corr})  gives rise to 
somewhat different height-related contributions to the 
Shapiro delay for the leading and trailing edges of the 
emission cone, if their pulse arrival times are 
measured separately.

\subsection{Aberration}
\label{sec:aberration}

Another time delay arises when the aberration from 
the velocity of the emitting region changes the 
spin phase of the pulsar at which its 
radio emission reaches the observer on
Earth. In a conventional approach, aberration is due
to $\Bp$, while in our case it is due to $\Bp+\Bs$.

As demonstrated in \cite{1976ApJ...207..574S}, DD, and
\cite{2006ApJ...641..438R} even in its conventional form 
(i.e. when $\Bs=0$) aberration leads not only to the 
overall homogeneous time delay of the  
profile (``longitudinal'' aberration delay) but also to the
distortion of the observed pulse shape. The latter
can be interpreted as the so-called ``latitudinal''
timing delay which has opposite signs for the different 
components of the pulse profile. The same result 
occurs in our case as well and we now  
consider these two effects separately.

\subsubsection{Longitudinal shift}
\label{sssect:long}

The change $\Delta \Phi$ of the pulsar spin phase that arises 
as a result of aberration can be expressed using simple 
geometric arguments as:
\begin{equation}
\Delta \Phi = \frac{\Dd\cdot(\v{s_p}\times\n0)}{|\v{s_p}\times \n0|^2}.
\label{eq:aberration}
\end{equation} 
Then, substituting in (\ref{eq:dom})-(\ref{eq:dsor}), 
the longitudinal timing delay is found to be
\ba
\Delta_A^{\rm long} = \frac{\Delta \Phi}{\Omega_p}
=\Delta_{A,0}^{\rm long}+\Delta_{A,h}^{\rm long}, 
\label{eq:ab_long_tot_sum}
\ea
where 
\ba
\Delta_{A,0}^{\rm long}=-\frac{\Bp\cdot(\sp\times\n0)}
{\Omega_p|\sp\times\n0|^2}
\label{eq:conv_ab_long}
\ea
is the conventional longitudinal aberration delay 
\cite{1976ApJ...207..574S} and
\begin{widetext}
\ba
&&\Delta_{A,h}^{\rm long} =\frac{\eO}{\Omega_p\sin^2\zeta}\nonumber\\
&&\times\left\{\sin^2\zeta(\Bp\cdot\x)(\r0\cdot\x)+\frac{{\bf \beta}_{o\perp}\cdot\m0}{3\cos\theta}
-\frac{\cos\zeta(3\cos^2\theta+1)^{1/2}}{3\cos\theta}(\sp\cdot\Bp)
-(\n0\cdot\Bp)\left[(\n0\cdot \r0)
-\cos\zeta(\sp\cdot\r0)\right]
\right\}.
\label{eq:h_ab_long}
\end{eqnarray}
\end{widetext}
Here we have 
retained only terms variable on the orbital
time scale (proportional to $\Bp$, i.e. orbital
and spin-orbital contributions).

Orbital aberration alone gives 
$\Delta_{A,0}^{\rm long}\sim \Bp \Omega_p^{-1} 
\simeq 3.3\mu s$. The height-related correction is smaller 
than $\Delta_{A,0}$ by $\epsilon_s$ so that 
$\Delta_{A,h}^{\rm long}\sim \Bp h c^{-1}=t_h$, where 
$t_h$ is a fiducial timescale, see equation (\ref{eq:t_h}). 
We discuss 
the measurability of the longitudinal aberration delay
in sec. \ref{sec:measurability}.

Note that unlike $\Delta_{A,0}^{\rm long}$ 
which is the same for all pulse components, 
the longitudinal height-related contribution 
$\Delta_{A,h}^{\rm long}$ contains terms with 
sign ambiguity (terms proportional to ${\bf m}_{0,x}$ 
or ${\bf r}_{0,x}$), meaning that in the case of finite 
$h$ even longitudinal aberration leads to pulse
distortion.

\subsubsection{Latitudinal shift}
\label{sssect:lat}

Aberration also changes the latitude $\zeta$ of 
vector $\doo$ with respect to the pulsar spin 
axis. This latitudinal shift is
\ba
\Delta \zeta_A = \frac{\Dd\cdot\sp}{|\sp\times\n0|}
=\Delta \zeta_{A,0}+\Delta \zeta_{A,h},
\ea
where
\ba
\Delta \zeta_{A,0}= -\frac{\Bp_\perp\cdot\sp}
{|\sp\times\n0|}
\label{eq:conv_ab_lat}
\ea
is the conventional latitudinal shift due to orbital
velocity \cite{1976ApJ...207..574S} and 
$\Delta \zeta_{A,h}=(\dsod\cdot\sp)/\sin\zeta$ 
is the contribution due to the finite height effects;
$\Delta \zeta_{A,h}$ is different for 
the leading and trailing pulse components because of 
the sign ambiguity in ${\bf m}_{0,x}$ 
and ${\bf r}_{0,x}$. 

Because of the pulsar beam shape, any shift in 
latitude results in the time delay of different
components of the pulse profile (see Fig. 4 of 
\citet{2006ApJ...641..438R} for an illustration).
Leading and trailing components corresponding
to the same emission cone are delayed by different 
amounts \cite{2006ApJ...641..438R}:
\ba
\Delta_A^{\rm lat} = \pm\frac{\Delta \zeta_A}
{\Omega_p\tan\chi_0}
=\Delta_{A,0}^{\rm lat}+\Delta_{A,h}^{\rm lat},
\label{eq:lat_dt}
\ea
where $\Delta_{A,0}^{\rm lat}$ is the latitudinal 
delay due to orbital aberration (see eqs. [22]-[23] in 
\cite{2006ApJ...641..438R}) and the contribution 
due to the finite height effects is 
\begin{widetext}
\ba
\Delta_{A,h}^{\rm lat}=
-\frac{\eO}{\Omega_p\tan\chi_0}
\left\{(\ro\cdot\x)\left[4\cos\zeta(\n0\cdot\Bp)-
(\sp\cdot\Bp)\right]+2\frac{(\Bp\cdot(\sp\times\r0))}
{\tan\zeta}
-\frac{\cos\zeta(3\cos^2\theta+1)^{1/2}}
{3\cos\theta}(\Bp\cdot\x)\right.\nonumber\\
+\left. \frac{({\bf \beta}_{o\perp}\cdot\m0)\left[
(\m0\cdot\n0)-\cos\zeta(\sp\cdot\m0)\right]}
{3\cos\theta\tan\zeta(\n0\cdot(\sp\times\m0))}\right\}.
\label{eq:h_ab_lat}
\ea
\end{widetext}
Here $\chi_0$ is the position 
angle of linear polarization given by 
\cite{1970Natur.225..612K}
\ba
\tan\chi_0=\frac{\sin\alpha\sin\Phi_0}{\cos\alpha\sin\zeta-
\cos\Phi_0\sin\alpha\cos\zeta}.
\label{eq:chi}
\ea
The magnitude of the latitudinal shift is in general 
(assuming $\tan\chi_0\sim 1$) of the same
order as the longitudinal one, i.e. one should expect 
$\Delta_{A,h}^{\rm lat}\sim t_h$. However,
in some cases, when the observer's line of sight
 cuts the emission cone very close to its edge,
$\tan\chi_0$ may become very small which increases
$\Delta_{A,h}^{\rm lat}\sim t_h/\tan\chi_0$, 
facilitating the detection of this delay 
\cite{2006ApJ...641..438R}.

The height-related contribution 
$\Delta_{A,h}^{\rm lat}$ contains both terms which are
the same for all pulse profile components and terms which 
have sign ambiguity. This is different from the 
orbital latitudinal delay, which has identical magnitude
but opposite signs for different crossings of the emission
cone. Based on the results of sec. \ref{sssect:long} and 
\ref{sssect:lat} we conclude that height-related corrections
change the simple separation of the effects of
longitudinal and latitudinal delays, i.e. a homogeneous shift
of the pulse profile in the former case and pulse distortion
in the latter. Both $\Delta_{A,0}^{\rm long}$
and $\Delta_{A,0}^{\rm lat}$ show a
homogeneous shift
as well as the distortion of the pulse profile.

\section{Measurability}
\label{sec:measurability}

Having calculated corrections to different time delays 
arising because of the finite height of emission region 
we can now ask whether these effects
are measurable. The magnitude of these corrections 
depends on the type and orbital parameters of the system
in question. In the double neutron star binary J0737-3039
\cite{2004Sci...303.1153L}
one finds $t_h\approx 0.33h_7~\mu$s, while the correction
to the Shapiro delay is about $0.06h_7$ ns, (neglecting 
factors which depend on angles $\alpha, \zeta, \eta$, etc.).
At the same time, the current TOA 
uncertainty in this system is about $18~\mu$s for a 30-s
integration at 820 MHz \cite{2006Sci...314...97K}, 
which makes height-dependent corrections 
unobservable on short timescales in this system,
despite its large value of $\beta_o=10^{-3}$.

Much better timing accuracies have been achieved in 
some NS-WD binaries, however small companion masses 
make $\beta_o$ significantly lower in these systems 
than in the DNS binaries. In particular, 
using the data from \cite{2006MNRAS.369.1502H} we find 
$t_h\approx 21h_7$ ns for J0437-4715 which has RMS 
timing residual $\sigma_{RMS}=200$ ns in 1-hour integration. 
For J1713+0747 we find $t_h\approx 12h_7$ ns 
($\sigma_{RMS}=125$ ns in 1-hour measurement), while  
for J1909-3744 $t_h\approx 30h_7$ ns
($\sigma_{RMS}=150$ ns in 1-hour integration). 
Thus, finite emission height effects are not currently 
observable in these systems. However, even more precise timing 
of these WD-NS binaries should allow future measurement 
of these effects by using many independent TOA measurements 
taken on longer time intervals. The
ongoing projects such as PPTA \cite{2006IAUJD..16E..66M},
 and the advent of SKA 
\cite{2004NewAR..48.1413C}, all of which dictate
understanding the systematic timing errors at the level
of $1$ ns, will enable the 
measurement of the finite height effects both
by lowering the timing errors in already known systems
and by discovering new binary pulsars which may be better
suited for this goal.

Another aspect of measurability has 
to do with the possible degeneracy between the
finite height corrections 
and the already known delays, which could arise if the 
orbital dependence of these corrections mimics the 
orbital variation of other delays. In particular, DD
have previously demonstrated that the longitudinal 
aberration delay\footnote{This expression can be obtained
by plugging (\ref{eq:betap}) into 
(\ref{eq:conv_ab_long}).} 
\ba
&& \Delta_{A,0}^{\rm long}=AS(\psi)+BC(\psi),
\label{eq:gen_form}\\
&& A=\frac{\tan\eta}{\cos i}B=-\frac{\Omega_b a_p\sin\eta}
{\Omega_p c\sqrt{1-e^2}\sin\zeta},
\label{eq:exp_aber}\\
&&S(\psi)\equiv\sin\psi+e\sin\omega,\nonumber\\
&&C(\psi)\equiv\cos\psi+e\cos\omega.
\label{eq:S&C}
\ea 
(with $\Omega_b=2\pi/P_b$) can be completely 
absorbed into the R\"omer delay by slightly rescaling the 
observed semimajor axis and eccentricity of the system
from their true values, e.g. 
\ba
\frac{a^{\rm obs}}{a^{\rm true}}=\frac{e^{\rm obs}}{e^{\rm true}}
= 1+\varepsilon_A,~~~~ 
\varepsilon_A=\frac{cA}{a_p\sin i},
\label{eq:absorb} 
\ea
where $a_p=aM_c/(M_p+M_c)$. This makes $\Delta_{A,0}^{\rm long}$ 
completely unobservable on an orbital timescale.

In fact, DD's argument can be generalized for {\it any} 
timing contribution that depends on the pulsar orbital 
phase as $XC(\psi)+YS(\psi)$, where $X$ and $Y$ are 
constant coefficients. The part of such a delay that 
is the same for all pulse components is then degenerate 
with the R\"omer delay. By the same argument, the sign-ambiguous delay
contributions (which are different for the leading and trailing 
pulse features and cause pulse profile distortions), 
are degenerate with the conventional 
latitudinal delay $\Delta_{A,0}^{\rm lat}$,
which has the orbital dependence analogous to 
(\ref{eq:gen_form}), see \citet{2006ApJ...641..438R}.

Using equations 
(\ref{eq:modroemer}), (\ref{eq:h_ab_long}), and 
(\ref{eq:h_ab_lat}) we see that $\Delta_{R,h}$,
$\Delta_{A,h}^{\rm long}$, and 
$\Delta_{A,h}^{\rm lat}$ are linear functions of 
$\Bp$, which in the coordinate system 
$(\x, \y, \z)$ is given by 
\ba
\Bp &=& \frac{\Omega_b a_p}{c\sqrt{1-e^2}}
\{\left[\cos i\cos\eta C(\psi)+\sin\eta S(\psi)\right]\x
\nonumber\\
& + &
\left[\cos i\sin\eta C(\psi)-\cos\eta S(\psi)\right]\y
\nonumber\\ & + &
\sin iC(\psi)~\z\}.
\label{eq:betap}
\ea
Thus, these finite 
height contributions are of the form 
(\ref{eq:gen_form}) with their own constant 
coefficients of order $t_h$. Capitalizing on the 
result of DD we can immediately say that all these
height-dependent contributions are degenerate 
with $\Delta_{R,0}$ and $\Delta_{A,0}^{\rm lat}$ 
and are unobservable on an orbital timescale. 
The existence of $\Delta_{R,h}$
and $\Delta_{A,h}^{\rm long}$
changes $A$ and $B$ in (\ref{eq:exp_aber}) to
$A+A_h$ and $B+B_h$, where $A_h/A\sim B_h/B\sim
\epsilon_s$ and the dependence of $A_h$
and $B_h$ on $\zeta,\eta, i$ can be explicitly 
written out using equations (\ref{eq:modroemer}), 
(\ref{eq:h_ab_long}), and (\ref{eq:h_ab_lat}). 
Based on this argument we can also claim that the height 
dependent delays affect the observed values of orbital 
parameters by changing $\varepsilon_A\to \varepsilon_A
+\varepsilon_h$, where $\varepsilon_h\sim 
ct_h/a_p= (h/a_p)\beta_o\sim 10^{-7}$. 

The orbital dependence of the Shapiro delay is 
unique enough to preclude $\Delta_{S,h}$ from 
being absorbed into either R\"omer or
latitudinal aberration delays. Nevertheless, we 
can look for a ``new'' orientation of the binary 
system characterized by the amended line of sight 
vector ${\bf \hat n}_{new} = \n0 + \delta\v{n}$, where
$\delta\v{n}$ is small and $\n0\cdot\delta\v{n}=0$.
Substituting $\n0={\bf \hat n}_{new} -\delta\v{n}$ into
(\ref{eq:shapiro}) and choosing 
\ba
\delta\v{n}=\frac{h}{|\v{R}_p-\v{R}_c|}
{\mathbf{\hat{r}}_{0\perp}}. 
\label{eq:dn}
\ea
we find that
\ba
\Delta_S = -\frac{R_g}{c}\log
\left[{\bf \hat n}_{new}\cdot(\v{R}_p-\v{R}_c) - 
|\v{R}_p-\v{R}_c|\right],
\label{eq:shapiro4}
\ea 
i.e. the expression for the Shapiro delay reduces 
to its standard form (\ref{eq:Shap_conv}). In general 
the small correction vector $\delta\v{n}$ varies 
with the pulsar orbital phase. However, 
the $\Delta_{S,0}$ attains its highest value near 
conjunction where the relative contribution of
the $\Delta_{S,h}$ to $\Delta_S$ is also 
largest and where $|\v{R}_p-\v{R}_c|$
is roughly constant. Based on this we conclude 
that $\Delta_{S,h}$ can be incorporated
into $\Delta_{S,0}$ by slight reorientation
of the binary system from $\n0$ to 
${\bf \hat n}_{new}$. This makes $\Delta_{S,h}$ 
unobservable on an orbital 
timescale due to its degeneracy with $\Delta_{S,0}$.

Despite the fact that timing contributions due to finite 
height cannot be separately measured on an orbital timescale,
one can hope to detect them on a longer geodetic precession 
timescale, analogous to the measurement of the conventional
aberration delay $\Delta_{A,0}^{\rm long}$ 
\cite[DD;][]{2004PhRvL..93n1101S,
2006ApJ...641..438R}. Geodetic precession 
changes a pulsar's spin orientation which, leads to periodic
variations of angles $\zeta$ and $\eta$, affecting both the 
conventional aberration coefficients $A$ and $B$ and 
the height-dependent timing contributions. According to
equation (\ref{eq:absorb}), this would lead to periodic
variation of the inferred values of the binary orbital
parameters (such as $a$ and $e$), and would allow one to
measure the combined effect of orbital aberration and
finite height effects. However, separating these two 
contributors to the secular timing effects from each other 
may not be easy, as they vary in a similar 
fashion. In addition, the contribution  to $\varepsilon$ 
from the finite emission height is rather small compared to 
$\varepsilon_A$. 
Additional information on the role of finite height effects 
compared to the orbital aberration may be provided 
by the pulse profile distortions on an orbital timescale 
(contributed both by $\Delta_{A,0}^{\rm lat}$ and
finite emission height effects), and by variation of the 
distortion pattern on the geodetic precession timescale.

Probably the easiest and most promising way of detecting the 
finite emission height effects is by using observations 
taken at different frequencies\footnote{This idea 
was suggested to us by Marten van Kerkwijk}. This method is 
based on the fact that the radio 
emission of different frequencies is produced at different
heights in pulsar magnetospheres \cite{2003A&A...397..969K, 
2004A&A...421..215M}, a manifestation of the so called
radius-to-frequency mapping (RFM) that will be discussed in 
more detail in sec. \ref{sect:var_height}. In this case 
$h=h(\nu)$ and by using timing measurements at different 
frequencies one would infer a different value of 
$a^{\rm obs}$ and $e^{\rm obs}$ at each frequency since
the height-dependent $\varepsilon_h$ is a 
function of $\nu$ as well. As the orbital aberration
is not frequency dependent, all variations of 
$a^{\rm obs}$ and $e^{\rm obs}$ that are frequency 
dependent must
be ascribed to the finite emission height effects.
By doing observations at sufficiently widely separated 
 frequencies, 
one can achieve corresponding height difference $\Delta h
\sim h$, in which case the relative offset between the
values of $a^{\rm obs}$ and $e^{\rm obs}$ at different 
frequencies would be of order $\varepsilon_h\sim 10^{-7}$.
This estimate suggests the level of accuracy to which 
$a^{\rm obs}$ and $e^{\rm obs}$ must be measured at a
set of frequencies in order for the height-dependent 
effects to be detected.

\section{Discussion}
\label{sect:discussion}

Our use of a specific geometric model of the radio emission 
(similar to RVM) does not imply that our results would be  
vitiated if pulsar beam has a non-circular shape. 
The assumption of a specific beam shape was necessary in
our case only to carry out an explicit calculation of various 
height related delays in terms of angles $\eta, \zeta, 
\theta$, etc. Clearly, even if pulsar has a non-circular 
beam, a finite emission height would still lead to  
aberration due to the pulsar spin and to the displacement
of the emission region from the pulsar center. Of course,
in this case we would not be able to predict exactly how 
the height-dependent timing signals scale with the pulsar 
orbital phase, which would complicate the interpretation 
of the secular evolution of these timing contributions 
on the geodetic precession timescale in terms of 
the pulsar spin orientation. However, even in this case 
one should still expect to see the difference in the 
observed orbital
parameters of the binary measured at different frequencies,
see the discussion in the end of sec. \ref{sec:measurability}.

We considered only the emission produced by a circular 
source (tied to a single flux surface) which has the form 
of two $\delta$-like spikes. This can be trivially 
generalized to the more realistic situation in which a
set of different flux surfaces characterized by different 
values of $\theta$ and $\rho$ generates emission of 
different intensities which is observed on Earth as a 
complex pulse profile. This is done by a directly summing 
such signals with the proper delay for each flux 
surface, reflecting not only the conventional time delays but
also the height-dependent contributions. Since the latter are 
explicit functions of $\theta, \rho$, etc. which are 
different for different flux surfaces, it is 
clear that the finite height effects would lead to the 
periodic distortions of the whole pulse profile 
(even within only the leading or trailing part of the
pulse profile) on an orbital timescale, as delays of 
different flux surfaces would vary differently. The 
relative temporal distortion of a pulse feature with 
width $\Delta \Phi$ due to the finite height effects 
should be at the level of $\sim \Delta \Phi \Omega_pt_h$
which can reach $\sim 10^{-4}$ for broad ($\Delta \Phi
\sim 0.1$) features in the profiles of the 
fastest ($P_p\approx 2$ ms) millisecond pulsars.

In this respect we would like to emphasize the importance 
of timing of the different components of the
pulse profile separately rather than just fitting single template to the
whole profile as is usually done. This procedure 
has been previously proposed for detecting  
orbital aberration on an orbital timescale 
\cite[DD;][]{2004PhRvL..93n1101S, 2006ApJ...641..438R}, 
and it is clear from the previous discussion that it may 
also be very useful for interpreting the timing effects
due to the finite emission height.

Our study was done in a monochromatic approximation while
real radio receivers always have finite bandwidth.  
We do not expect this to affect our conclusions in any 
significant way, as the broadband pulse profile 
can be obtained by simply convolving the monochromatic pulse
shape of the pulsar with the window function of the receiver. 
As long as one uses non-overlapping (even if broad) bands 
one would still see the difference in the observed
orbital parameters of the binary in different bands, which
would in principle be a clear signature of the finite 
emission height effects.

\subsection{Delays due to frequency dependent Emission Height}
\label{sect:var_height}

Our calculations have explicitly assumed the height of the
emission region $h$ to be constant. In reality, observations 
\cite{2001ApJ...555...31G, 2003A&A...397..969K, 
2004A&A...421..215M} find evidence for a radius-to-frequency
mapping in which radio photons of different 
frequencies are emitted at different heights, so that 
$h=h(\nu)$. 

It is then easy to see that the orbital motion of pulsar should
give rise to additional time delays proportional to 
$dh/d\nu$. Indeed, because of the 
orbital Doppler shift, a monochromatic receiver detects 
radio photons which were emitted at slightly different
frequency ($\Delta\nu/\nu\sim \beta_o\sim 10^{-3}$)  
than in the stationary case. The $h(\nu)$ relation
then implies that the height of emission also changes 
periodically to reflect this Doppler shift as\footnote{The linear
character of the relation between $\Delta h$ and $\Delta\nu$ 
is not affected by the variation of the gravitational redshift 
in the field of pulsar and the change of spin velocity of
the emission region. The coefficient in the relation changes
because of these additional effects but only at the 
level $R_g/h$ and $\beta_s$ correspondingly, 
where $R_g$ is the Schwarzschild radius of the pulsar.} 
$\Delta h/h\sim \Delta\nu/\nu\sim \beta_o$ (observations 
suggest that $d\ln h/d\ln \nu$ is roughly
constant). This clearly translates into another correction
to the R\"omer delay which is about $\Delta h/c\sim \beta_o h/c$,
i.e. has the same magnitude as $\Delta_{R,h}$ and all 
other height-dependent delays calculated in this paper.

Another effect arises because of the direct relationship 
between the distance from the pulsar center $h$ and the
magnetic latitude angle $\theta$ for a given flux surface:
\begin{equation}
h(\theta) = r_e\sin^2\theta,
\label{eq:dip}
\end{equation}
where $r_e$ is the equatorial radius of the flux surface, 
assuming that the dipolar geometry holds even beyond the 
light cylinder. The same Doppler effect that causes variations
in $h$ also causes the emission 
cone to widen at some orbital phases and narrow at others
with the amplitude
\ba
\Delta \theta=\frac{\tan\theta}{2}\frac{\Delta h}{h},
\label{eq:dtheta} 
\ea
which causes pulse profile distortions. Because of 
that the direction of emission also changes by
\ba
\Delta_\theta \doo=\frac{\partial\doo}{\partial\theta}
\Delta \theta=-\frac{3\sin\theta(\ro+\cos\theta\mo)}
{(3\cos^2\theta+1)^{3/2}}\Delta\theta,
\label{eq:dd}
\ea
see equation (\ref{eq:d_through_r_m}).
Similar to the latitudinal aberration delay 
\cite{2006ApJ...641..438R} such a distortion can 
be interpreted in terms of the corresponding time delay
$\sim|\Delta_\theta \doo|\Omega_p^{-1}$ which, according
to (\ref{eq:dtheta}) and (\ref{eq:dd}), is of order 
$\Omega_p^{-1}\beta_o\sin^2\theta=\beta_o h/(\Omega_p r_e)$.
But since pulsar radio emission is produced in the open field
line region, $r_e\gtrsim r_{LC}=c/\Omega_p$, this 
additional delay is $\lesssim \beta_o h/c$ -- the same order
of magnitude as all other height-dependent delays.

There are two reasons why we have not explicitly computed
timing effects arising from RFM despite the fact that 
they are non-negligible compared to the contributions 
which we have already calculated.
First, all of these effects are proportional to $d\ln h/d\ln \nu$ 
while the height to frequency dependence is not very strong,
$|d\ln h/d\ln \nu|\approx 0.2-0.3$ \cite{2003A&A...397..969K}.
Second, the introduction of these terms does not change the 
qualitative picture of the finite emission height effects 
on timing and our conclusions about their measurability,
but it does make calculations much more cumbersome.
Because of this, we decided to leave the careful calculation 
of these effects for a future work.

\section{Summary}
\label{sec:discussion}

We have extended the standard calculation of timing delays in
binary millisecond pulsar systems by taking into account the fact that
the emission region does not spatially coincide with the
physical center of the pulsar but is located at some height
(tens to thousands of kilometers) above the neutron star
surface. This very natural extension has two consequences. First,
such a  displacement changes the propagation (R\"omer
and Shapiro) delays. Second, the large spin velocity of the 
high-altitude emission region affects the aberration of the
radio signal and contributes to the aberration delay. We 
have calculated the height-dependent corrections to these
delays in the framework of a specific model of the pulsar 
emission geometry. The typical magnitude of corrections 
$(h/c)\beta_o$ is of order several hundred nanoseconds 
in compact DNS systems (which is at the current 
precision level of a few most accurately timed millisecond 
pulsars in NS-WD binaries), 
and should be measurable in the future by projects like
PPTA and facilities like SKA. Although these timing corrections 
are degenerate with other delays (R\"omer and aberration), they
can be detected using multi-frequency 
observations and exploiting the RFM. Detection
of these effects could yield a new measure of the 
emission height in pulsars, which would provide a 
consistency check on the estimates using other methods.
Although our calculation assumes a specific model of
the emission region geometry (that of a circular emission 
cone in a dipolar magnetic field), the main results 
of our analysis are qualitatively independent of that model.

\acknowledgments

We are grateful to Marten van Kerkwijk for illuminating
discussions and useful suggestions. The financial support 
for this work was provided by the Canada Research Chairs
program and NSERC.

\bibliography{ms}

\appendix
\section{Corrections to vectors $\mo$ and $\ro$}

Here we compute changes in the vectors $\ro$ and $\mo$ 
that result from 
the orbital and spin motion of the emission region. The 
corrections we are looking for are small, typical orders 
are $\epsilon_{so}\ll\epsilon_{o}\ll\epsilon_s\ll 1$ for
the spin-orbital $\Delta_{so}$, orbital $\Delta_{o}$, spin
$\Delta_{s}$, and static terms correspondingly (the relative 
size of each of these terms is given in equations 
(\ref{eq:eps_orb})-(\ref{eq:eps_so})).  Thus, we
can employ perturbation theory to compute these corrections
by consecutively going to different orders in these 
small parameters. 

In the static limit we neglect any motion of the emission 
region in which case equations (\ref{eq:mag_spin}), 
(\ref{eq:d_through_r_m}), (\ref{eq:m_r}),
and (\ref{eq:master_eq}) supplemented with the condition
$|\mo|=1$  result in equations (\ref{eq:m0}) and 
(\ref{eq:r0}).

\subsection{Contributions due to pulsar spin}

In the next order we retain only terms proportional 
to the spin velocity $\Bs c$ of the emission region,
where $\Bs=\epsilon_s(\sp\times\r0)$. This motion aberrates the
vector $\doo$, producing an additional contribution 
$\dsd$  to $\d0=\n0$ so that instead 
of (\ref{eq:master_eq}) we now have
\begin{equation}
\doo = \d0 + \dsd = \n0 - \Bs_\perp
\label{eq:master_spin}
\end{equation}
as the condition for the pulse to be emitted towards the
observer. For any vector ${\bf A}$, we denote its component 
perpendicular to $\n0$ as ${\bf A}_\perp \equiv {\bf A} 
- ({\bf A}\cdot\n0)\n0$. 

From the definition (\ref{eq:d_through_r_m}) we find that
(assuming constant $\theta$, i.e. no RFM, see sec. 
\ref{sect:var_height}):
\ba
\dsd =\frac{3\cos\theta \dsr - \dsm}{(3\cos^2\theta + 1)^{1/2}},
\label{eq:dd_dr_dm_spin}
\ea
while equations (\ref{eq:mag_spin}), (\ref{eq:m_r}) and 
condition $|\mo|=1$ lead to three more constraints:
\begin{eqnarray}
\dsm\cdot\m0 &=& 0,\\
\nonumber\dsm\cdot\sp &=& 0,\\
\nonumber\dsm\cdot\r0 &=& -\dsr\cdot\m0.
\label{eq:3eq_spin}
\end{eqnarray}
Combining equations (\ref{eq:master_spin}), 
(\ref{eq:dd_dr_dm_spin}), and (\ref{eq:3eq_spin}) we obtain six
equations for six unknowns (components of $\dsm$ and $\dsr$),
which can be easily solved:
\begin{eqnarray}\label{eq:dmdr}
\dsm &=& \frac{(\Bs_\perp\cdot\m0)(\sp\times\m0)}
{\n0\cdot(\sp\times \m0)},
\label{eq:dsm}\\
\dsr &=& \frac{(\Bs_\perp\cdot\m0)(\sp\times\m0)}
{3\cos\theta\n0\cdot(\sp\times \m0)} \nonumber\\
& - & \frac{(3\cos^2\theta + 1)^{1/2}}
{3\cos\theta}\Bs_\perp,
\label{eq:dsr}
\end{eqnarray}
providing us with the sought corrections due to the spin motion 
of the emission region. Note that although $\epsilon_s^2$ 
can easily exceed $\epsilon_o$ we do not need to go to higher
order in $\epsilon_s$ because the spin contributions $\Delta_s$ 
do not vary with the orbital phase of the pulsar and thus 
are undetectable by timing. On the other hand, despite the fact that
the constant $\dsm$ and $\dsr$ given by (\ref{eq:dsm}) and 
(\ref{eq:dsr}) do not affect timing directly, we need to 
compute them in order to calculate the spin-orbital 
corrections in the next section.

\subsection{Corrections due to orbital motion and 
spin-orbit coupling}

In general the aberration of vector $\doo$ is caused by the combination
of both orbital and spin motions of the emission region.
This gives rise to additional orbital contributions $\Delta_o$ (which
exist even if pulsar is not spinning or if $h=0$, an assumption used by DD), and spin-orbital contributions
$\Delta_{so}$ to vectors $\mo$ and $\ro$, related to corresponding
$\Delta_o {\bf d}$ and $\Delta_{so} {\bf d}$ via equations analogous
to (\ref{eq:dd_dr_dm_spin}).

Instead of equation (\ref{eq:master_spin}) we now have
\ba
\doo & = & \d0+\dsd + \dod+\dsod \nonumber\\
& = & \n0 - \Bp + 
\doo(\Bp\cdot\doo) - \Bs+\doo(\Bs\cdot\doo).
\label{eq:totald}
\ea
Substituting $\doo=\d0+\dsd+\Delta_o {\bf d}$
in the right-hand side of this equation (in doing that 
we ignore the spin-orbital term $\dsod$ as it would lead to higher order 
corrections) one obtains
\ba
\Delta_o {\bf d} &=& -\Bp_\perp 
\label{eq:master_o}\\
\Delta_{so} {\bf d} &=& 
-\epsilon_s[\Bp_\perp(\n0\cdot(\sp\times\r0)) + (\n0\cdot\Bp)(\sp\times\r0)_\perp 
\nonumber\\
&+&  2\n0(\Bp_\perp\cdot(\sp\times\r0)) + (\sp\times\dor)_\perp].
\label{eq:master_so}
\ea
Finally, instead of (\ref{eq:3eq_spin}) we have the following 
relations:
\begin{eqnarray}
\label{eq:spin11}
&& (\dom+\dsom) \cdot\sp = 0,\\
&& (\dom+\dsom)\cdot(\r0+\dsr) = \nonumber\\
&& -(\m0+\dsm)\cdot(\dor+\dsor),\\
&& (\dom+\dsom)\cdot(\m0+\dsm) = 0,
\end{eqnarray}
which provide us with enough information to solve for 
$\dor, \dom, \dsor, \dsom$ by considering separately 
terms of order $\epsilon_o$ and $\epsilon_{so}$. 

After tedious but straightforward calculations we find 
\begin{eqnarray}
\dom &=& \frac{\sp\times\m0}{\n0\cdot(\sp\times\m0)}
(\Bp_\perp\cdot\m0)
\label{eq:dom}\\
\dor &=& \frac{\sp\times\m0}{3\cos\theta 
\n0\cdot(\sp\times\m0)}(\Bp_\perp\cdot\m0)\nonumber \\
 &-& \frac{(3\cos^2\theta + 1)^{1/2}}{3\cos\theta}\Bp_\perp
\label{eq:dor}.
\end{eqnarray}
for the orbital terms and 
\begin{widetext}
\begin{eqnarray}
\dsod &=& - \eO
\Bigg[\Bp_\perp(\n0\cdot(\sp\times\r0)) +2\n0(\Bp_\perp\cdot(\sp\times\r0))
- \frac{(3\cos^2\theta + 1)^{1/2}}
{3\cos\theta}(\sp\times\Bp_\perp)_\perp \nonumber\\
&+& (\n0\cdot\Bp)(\sp\times\r0)_\perp
+\frac{(\sp\times(\sp\times\m0))_\perp}{3\cos\theta}\frac{\Bp_\perp\cdot\m0}{\n0\cdot(\sp\times\m0)}\Bigg]
 \label{eq:dsor}
\end{eqnarray}
\end{widetext}
for contributions to $\doo$ resulting from the 
spin-orbital coupling (we do not separately display
expressions for $\dsor$ and $\dsom$ as they are very 
cumbersome and are not used separately in the timing
calculations anyway).
These expressions are then used in sec. \ref{sec:timing} 
to compute the height-related corrections to different 
time delays.

\end{document}